\documentclass[twocolumn,prl,amsmath,amssymb,showpacs]{revtex4}
\usepackage{graphicx}
\usepackage{dcolumn}
\usepackage{bm}
\usepackage{latexsym}
\usepackage{amstext}
\usepackage{amssymb}
\usepackage{amsxtra}
\usepackage[english]{babel}

\newcommand{\op}{\omega_\perp}

\begin{document}
\title{Observation of Phase Defects in Quasi-2D Bose-Einstein Condensates}
\author{Sabine Stock, Zoran Hadzibabic, Baptiste Battelier, Marc Cheneau, and Jean Dalibard}
\affiliation{Laboratoire Kastler Brossel$^{*}$, 24 rue Lhomond,
75005 Paris, France}
\date{\today}

\begin{abstract}

We have observed phase defects in quasi-2D Bose-Einstein
condensates close to the condensation temperature. Either a single
or several equally spaced condensates are produced by selectively
evaporating the sites of a 1D optical lattice. When several clouds
are released from the lattice and allowed to overlap, dislocation
lines in the interference patterns reveal nontrivial phase
defects.

\end{abstract}


 \maketitle

Low dimensional bosonic systems have very different coherence
properties than their three dimensional (3D) counterparts. In a
spatially uniform one dimensional (1D) system, a Bose-Einstein
condensate (BEC) cannot exist even at zero temperature. In two
dimensions (2D) a BEC exists at zero temperature, but phase
fluctuations destroy the long range order at any finite
temperature. At low temperatures the system is superfluid and the
phase fluctuations can be described as bound vortex-antivortex
pairs. At the Kosterlitz-Thouless (KT) transition
temperature~\cite{kost73, bish78, safo98} the unbinding of the
pairs becomes favorable and the system enters the normal state.

In recent years, great efforts have been made to study the effects
of reduced dimensionality in trapped atomic
gases~\cite{pric04note}. In both 1D and 2D, the density of states
in a harmonic trap allows for Bose-Einstein condensation at finite
temperature. In contrast to 1D and elongated 3D
systems~\cite{schr01, gorl01dim, dett01, shva02, rich03, tolr04,
stof04, pare04, kino04}, the coherence properties of 2D atomic
BECs have so far been explored only
theoretically~\cite{petr00BEC2D, kaga00, ande02, simu04}. In
previous experiments, quasi-2D BECs~\cite{gorl01dim, schw04,
rych04, smit05} or ultracold clouds~\cite{colo04} were produced in
specially designed ``pancake" trapping potentials. The sites of a
1D optical lattice usually also fulfill the criteria for 2D
trapping~\cite{orze01, burg02, hadz04, kohl05}; the difficulty in
these systems is to suppress tunneling between the sites, and to
address or study them independently~\cite{ott04, schr04}.

In this Letter, we report the production of an array of
individually addressable quasi-2D BECs. By selectively evaporating
the atoms from the sites of a 1D optical lattice, we can produce
either a single or several equally spaced condensates. The
distinct advantage of this approach is that it opens the
possibility to study the phase structures in quasi-2D BECs
interferometrically. We have observed interference patterns which
clearly reveal the presence of phase defects in condensates close
to the ideal gas Bose-Einstein condensation temperature. We
discuss the possible underlying phase configurations.

\begin{figure}
\vspace{1mm} \centerline{\includegraphics[width=8.5cm]{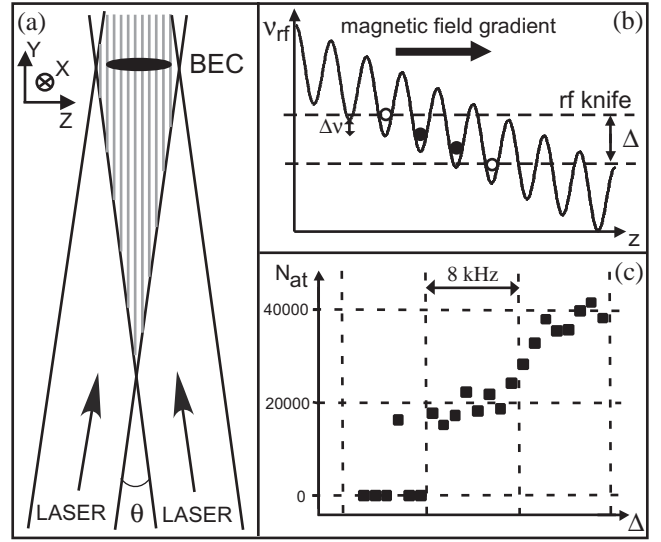}}
\caption{An array of individually addressable quasi-2D BECs. (a)~A
1D optical lattice splits a cigar-shaped 3D condensate into 15 -
30 independent quasi-2D BECs. (b)~A magnetic field gradient along
the lattice axis allows us to selectively address the sites by an
rf field. We evaporate the atoms from all the sites except those
within a frequency gap $\Delta$. (c)~Steps in the BEC atom number
$N_{\rm at}$ as a function of $\Delta$, corresponding to 0, 1 and
2 sites spared from evaporation. Each data point represents a
single measurement.} \label{fig:fig1} \vspace{-5mm}
\end{figure}

Our experiments start with an almost pure $^{87}$Rb condensate
with $4\times10^5$ atoms in the $F=m_F=2$ hyperfine state,
produced by radio-frequency (rf) evaporation in a cylindrically
symmetric Ioffe-Pritchard (IP) magnetic trap.  The trapping
frequencies are $\omega_z/2\pi=12\,$Hz axially, and
$\op/2\pi=106\,$Hz radially, leading to cigar-shaped condensates
with a Thomas-Fermi length of $90\,\mu$m and a diameter of
$10\,\mu$m.

After creation of the BEC we ramp up the periodic potential of a
1D optical lattice, which splits the 3D condensate into an array
of independent quasi-2D BECs (see Fig.~\ref{fig:fig1}(a)
and~\cite{hadz04}). The lattice is superimposed on the magnetic
trapping potential along the long axis ($z$) of the cigar. Two
horizontal laser beams of wavelength $\lambda=532\,$nm intersect
at a small angle $\theta$ to create a standing wave with a period
of $d=\lambda/[2\sin(\theta/2)]$. The blue-detuned laser light
creates a repulsive potential for the atoms, which accumulate at
the nodes of the standing wave, with the radial confinement being
provided by the magnetic potential. Along $z$, the lattice
potential has the shape $V(z)=V_{0}\cos^{2}(\pi z/d)$, with $V_0/h
\approx 50\,$kHz.

For the work presented here we have used two lattice periods,
$d_1=2.7\,\mu$m and $d_2=5.1\,\mu$m. The respective oscillation
frequencies along $z$ are $\omega_1/2\pi= 4.0\,$kHz and
$\omega_2/2\pi= 2.1\,$kHz. At the end of the experimental cycle
(described below), the BEC atom numbers in the most populated,
central sites are $N_1\approx 10^{4}$ and $N_2\approx 2\times
10^{4}$. We numerically solve the Gross-Pitaevskii equation to get
the corresponding chemical potentials $\mu_1/h=2.2\,$kHz and
$\mu_2/h=2.5\,$kHz, where the $\hbar\omega_{1,2}/2$ zero-point
offset is suppressed in our definition of $\mu$. In the smooth
crossover from 3D to 2D, the condensates in the shorter period
lattice are thus well in the 2D regime with
$\mu_1/(\hbar\omega_1)= 0.6$, while for the clouds in the longer
period lattice this ratio is $1.2$.

Since the radial trapping is purely magnetic, we can remove atoms
from the lattice by rf induced spin-flips to untrapped Zeeman
states. In order to address the lattice sites selectively, we
apply a magnetic field gradient $b'$ along $z$~\cite{schr04, QP}.
This creates an energy gradient along the lattice direction, and
splits the resonant frequencies for evaporation of atoms from two
neighboring sites by $\Delta\nu_{1,2}=\mu_{\textrm{B}}b'd_{1,2}/(2
h)$, where $\mu_{\textrm{B}}$ is the Bohr magneton
(Fig.~\ref{fig:fig1}(b)). We use gradients up to 26$\,$G/cm,
corresponding to $\Delta\nu_1=5\,$kHz and $\Delta\nu_2=9\,$kHz.
These splittings are larger than the chemical potentials
$\mu_{1,2}$, and the rf Rabi frequency ($\approx 2\,$kHz). The
lattice sites can thus be addressed individually.

The experimental routine to produce an adjustable number of
condensates starts with a slow, 200$\,$ms ramp-up of the gradient
$b'$. As illustrated in Fig.~\ref{fig:fig1}(b), we then evaporate
the atoms from both ends of the cigar, sparing only the central
sites within a variable rf frequency gap $\Delta$. We perform this
evaporation in 100$\,$ms, switch off the rf field, and ramp $b'$
back to zero in another 200$\,$ms~\cite{sag}. During this time,
some heating of the remaining clouds occurs, and they reach a
temperature slightly below the condensation temperature, as we
discuss in detail below.

To verify that we can address the lattice sites individually, we
measure the total number of condensed atoms left in the trap as a
function of $\Delta$. An example of such a plot is shown in
Fig.~\ref{fig:fig1}(c) for $d_2=5.1\,\mu$m and $b'=22\,$G/cm. The
magnetic and optical trap were switched off simultaneously and the
atomic density distribution was recorded by absorption imaging
along $z$ after 18$\,$ms of time-of-flight (TOF) expansion. The
atom number increases in steps of $N_2=2\times 10^{4}$ every
$8\,$kHz, in agreement with the expected $\Delta\nu_2$. We see
three clear plateaus corresponding to 0, 1 and 2 sites spared from
evaporation. For the shorter lattice period the frequency
splitting is comparable to the chemical potential. This results in
some rounding off of the steps, but the plateaus remain visible.


\begin{figure}
\centerline{\includegraphics[width=8.5cm]{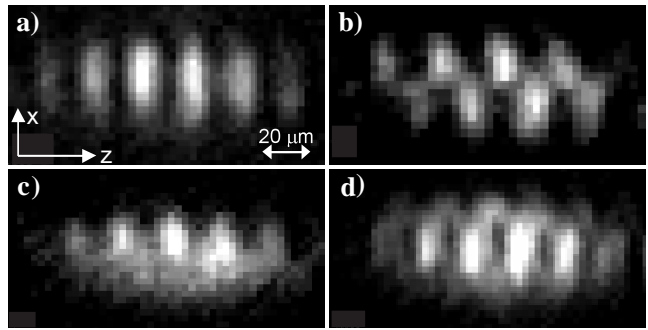}} \vspace{-1mm}
\caption{Phase defects in quasi-2D condensates. Interference of
four (a-c) and seven (d) independent BECs is observed 12$\,$ms
after release from the $2.7\,\mu$m period lattice. Dislocation
lines in the interference patterns (b-d) reveal the presence of
phase defects in quasi-2D condensates.} \label{fig:fig2}
\vspace{-5mm}
\end{figure}


In the first set of experiments, we have characterized the free
expansion of a single quasi-2D BEC~\cite{hech05}. The clouds were
released from the $5.1\,\mu$m period lattice and imaged after up
to 18$\,$ms of TOF. We extract the axial ($l$) and the radial
($w$) rms size of the cloud from gaussian fits to the density
distribution. As might be expected, the observed expansion is
predominantly one-dimensional, along the axial direction. For
short expansion times, $t \leq 3\,$ms, the apparent axial size is
limited by our imaging resolution, but for longer times it follows
the linear scaling $l=vt$, with $v=2.7\,$mm/s. This value is
comparable to the calculated velocity in the harmonic oscillator
ground state along $z$, $v_g=\sqrt{\hbar\omega_2/(2m)}=2.1\,$mm/s,
where $m$ is the atomic mass. We find that the radial expansion
can be described by the empirical law
$w=w_{0}\sqrt{1+(t/t_{0})^{2}}$, with $w_{0}=4.4\,\mu$m and
$t_{0}=5.7\,$ms. The same law describes the radial expansion of a
cigar-shaped 3D condensate, with
$t_{0}=\omega_{\perp}^{-1}$~\cite{cast96}, where
$\omega_{\perp}^{-1}=1.5\,$ms for our trap. The radial expansion
of our 2D gas is slower by a factor of $\approx 4$ compared to the
3D case, because the fast axial expansion results in an almost
sudden ($\omega_2^{-1}=76\,\mu$s) decrease of the atomic density,
and only a small fraction of the interaction energy is converted
into radial velocity.

In the second set of experiments, we have studied interference of
independent quasi-2D BECs. Between two and eight clouds were
released from the 2.7$\,\mu$m period lattice and allowed to expand
and overlap~\cite{periods}. The resulting interference patterns
were recorded by absorption imaging along the radial direction
$y$. Due to the finite imaging resolution we observe only the
first harmonic of the interference pattern with period
$ht/(md_{1})$. Each image is thus the incoherent sum of the
pairwise interferences of nearest-neighbor condensates.

Interference of equally spaced, independent BECs produces straight
interference fringes (Fig.~\ref{fig:fig2}(a)) as long as each BEC
has a spatially uniform phase~\cite{andr97int, hadz04}. The main
result of this paper is the observation of topologically different
patterns, which reveal the presence of phase defects in quasi-2D
condensates.  Striking examples are ``zipper" patterns
(Fig.~\ref{fig:fig2}(b)), where the fringe phase changes abruptly
by $\pi$ across a dislocation line parallel to $z$. On both sides
of the dislocation, the fringe contrast is as high as in
Fig.~\ref{fig:fig2}(a). We also observe ``comb" patterns
(Fig.~\ref{fig:fig2}(c)), which show a dislocation with high
fringe contrast on one side of the line, and vanishing on the
other. Finally, we sometimes see ``braid" structures with two
dislocation lines (Fig.~\ref{fig:fig2}(d)). Single dislocations
(zippers and combs) are clearly visible in about 15$\%$ of 200
experiments with four interfering clouds~\cite{2D_lowcontrast}. To
verify that the occurrence of defects is an equilibrium property
of the system, we have checked that dislocations are still
observed when holding the clouds in the lattice for 500$\,$ms
after ramping down the gradient $b'$.

\begin{figure}
\vspace{1mm}
\centerline{\includegraphics[width=8.5cm]{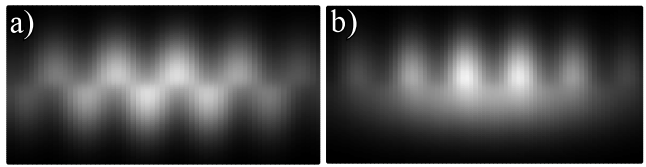}}%
\vskip -2mm \caption{Examples of numerical simulations of two (a)
and four (b) interfering condensates. In both cases one randomly
chosen BEC has a phase factor $e^{i\varphi}$ corresponding to a
centered vortex, and the others have randomly chosen uniform
phases. For simplicity, we model the clouds as gaussian wave
packets and neglect interactions during the expansion. The images
are convolved with a gaussian of $4\,\mu$m rms width to simulate
the finite imaging resolution.} \label{fig:fig3} \vspace{-5mm}
\end{figure}

The simplest phase configuration which can produce a sharp
dislocation line is a single vortex in one of the condensates (see
also~\cite{bold98, temp98, cast99, inou01vort, chev01}). In the
case of two interfering BECs, one can show that a centered vortex
always leads to a zipper pattern (see a simulation of the expected
pattern in Fig.~\ref{fig:fig3} (a)). The zipper is indeed the only
type of dislocation we clearly observe with two clouds. When more
than two BECs interfere, the presence of a single vortex can
result both in a zipper and in a comb pattern, depending on the
phases of the other condensates. In Fig.~\ref{fig:fig3} (b) we
show a numerical simulation with four BECs leading to a comb.
Increasing the number of interfering BECs enhances the probability
that some of them contain defects~\cite{2D_braids}, but the
interpretation of images also becomes increasingly difficult.
Further, for a large number of clouds, a single defect will not
produce a clear dislocation line in the first harmonic of the
interference pattern, because it affects only the interference
with the two neighboring BECs. Already with four clouds, only half
of 100 simulations with a vortex show clear zipper- or comb-type
dislocations. The other half shows weaker dislocations which are
not easily distinguishable from straight interference fringes.

Despite the agreement between simulations involving a vortex and
the observed patterns, we point out that it is in general not
possible to unambiguously deduce the underlying phase
configuration from an interference image. For example, a
dislocation line could also come from a dark soliton, where the
phase of one of the BECs changes by $\pi$ across a line parallel
to the imaging axis. In future experiments simultaneous imaging
along a second radial direction could allow us to discriminate
between different possible phase structures leading to the
observed interference patterns.

So far we could not observe a clear signature of vortices in
images of single condensates taken along the axial direction $z$.
We suspect that this is difficult because of the expansion
properties of a 2D gas. Rotating 3D BECs, in which vortices are
readily detected after TOF~\cite{madi00}, expand mostly radially,
while our clouds expand mostly axially. Therefore, any small
misalignment with the imaging axis will significantly reduce the
contrast. Interferometric detection along a radial direction
offers a fundamentally superior signal, because a localized defect
affects the appearance of the whole image.

It is important to assess the temperature of the clouds in which
the observed phase defects appear. Precise thermometry at the end
of the experimental cycle is difficult, because the thermal cloud
is very dilute. However, we can estimate lower and upper bounds
for the temperature. During the 500$\,$ms selective evaporation
routine, the clouds are heated due to three-body recombination,
and the only constant source of cooling is the finite lattice
depth; atoms with an energy larger than $V_{0}$ are accelerated
away by the magnetic field gradient (Fig.~\ref{fig:fig1}(b)).
Assuming the largest realistic evaporation parameter $\eta =
V_{0}/(kT)=10$, we get a lower bound for the temperature $T_{\rm
min} \approx 250\,$nK. To get an upper bound we note that at the
beginning of the experiment the condensed fraction is certainly
above 50$\%$. During the experimental cycle the number of
condensed atoms in the remaining sites drops by a factor of
$\approx 2$. This means that, even if we neglect losses in the
total atom number, the final condensed fraction cannot be less
than 25$\%$. Using the measured number of condensed atoms and
integrating the Bose distribution over the density of states in
the lattice, we get $T_{\rm max} \approx 500\,$nK. Since the
number of thermal atoms is different at $T_{\rm min}$ and $T_{\rm
max}$, the two bounds correspond to different condensation
temperatures $T_c$, and the estimated temperature range is more
clearly expressed as $0.7 \leq T/T_c \leq 0.9$. In this
temperature range $k T \gtrsim \hbar \omega_{1,2}$, so the thermal
clouds are not fully in the 2D regime.

The fact that the clouds are close to $T_c$ is probably essential
for the understanding of our observations, and a systematic
temperature study will be the subject of future work. The
probability for a thermal excitation of the system into a vortex
state is $\propto e^{-F/kT}$, where $F=E-TS$ is the free energy
associated with the excitation, $E$ the energy, and $S$ the
entropy. Here we estimate the conditions for $F/kT$ to be of order
unity. For a vortex in the center of the condensate, $E \sim N
\left[(\hbar \op)^2/\mu
\right]\ln\left(R/\xi\right)$~\cite{cast99}, where $N$ is the BEC
atom number, $R$ the size of the condensate, and
$\xi=\hbar/\sqrt{2m\mu}$ the size of the vortex core.
Equivalently, $E/(k T) \sim (1/2)\, n_0 \lambda^2 \,
\ln\left(R/\xi\right)$, where $n_0$ is the peak 2D atom density
and $\lambda$ is the thermal wavelength $h/\sqrt{2\pi mkT}$. The
number of distinguishable positions for a straight vortex of size
$\xi$ in a region of size $R$ is $\sim R^2/\xi^2$, and the
associated entropy is $S/k\sim 2 \ln\left(R/\xi\right)$. In this
estimate $F/kT\propto \left(n_0 \lambda^2 - 4\right)$ vanishes for
$n_0 \lambda^2 = 4$. In our experiment $n_0 \lambda^2\sim 10-20$
is a few times higher than this value. However already this crude
agreement suggests that the thermal excitation of vortices might
be possible in our system.

Thermal excitation of a tightly bound vortex-antivortex
pair~\cite{simu04} is more likely than that of a single vortex. In
that case the entropy is comparable and the energy is typically
lower by the logarithmic factor $\ln\left(R/\xi\right)$, in our
case $\sim 4$. These pairs are difficult to detect with our
interferometric scheme since they create only small phase slips in
the fringe pattern. However they can play a significant role by
screening the velocity field of a single vortex, thus lowering its
energy and making its appearance more likely~\cite{kost73}.

The fact that $\ln\left(R/\xi\right)$ is not large compared to 1
underlines the mesoscopic nature of our system. In a homogeneous
2D system with $R \rightarrow \infty$, both the energy and the
entropy of a free vortex diverge as $\ln(R)$, and the two
contributions to the free energy cancel at the KT transition
temperature $T_{\rm KT}$. Below $T_{\rm KT}$ only
vortex-antivortex pairs are present, while above $T_{\rm KT}$ a
large density of free vortices appears and suppresses
superfluidity. In our case, we expect this phase transition to be
replaced by a gradual increase of the average number of free
vortices with temperature. For $F \sim kT$, the vortex number can
show large fluctuations and two condensates produced under
identical experimental conditions can have qualitatively different
wave functions.

In conclusion, by selectively addressing individual sites of a 1D
lattice, we have produced both a single and several equally spaced
quasi-2D BECs. We have characterized the free expansion of a
single BEC, and have interferometrically observed clear evidence
for the presence of phase defects in about 10$\%$ of eight hundred
condensates. While our observations can be explained by the
presence of thermally excited vortices in the system, this does
not exclude other scenarios and we hope that our experiments will
stimulate further theoretical work.

\acknowledgments We thank the ENS ``cold atoms" group for useful
discussions. S.S. acknowledges support from the Studien\-stiftung
des deutschen Volkes and the DAAD, and Z.H. from a Chateaubriand
grant and the EU (Contract No. MIF1-CT-2005-00793). This work is
partially supported by CNRS, R\'egion Ile de France, and ACI
Nanoscience.


\begin{thebibliography}{39}
\bibitem[{QP()}]{QP}
\bibinfo{note}{The gradient is produced by a pair of coils in anti-Helmholtz
  configuration. Their quadrupole field splits the radial trap frequencies
  $\omega_{x,y}$ by $20\%$ for $b'=26\,$G/cm.}
\bibitem[{sag()}]{sag}
\bibinfo{note}{To improve addressability we reduce the gravitational sag
to $6\,\mu$m for the evaporation of individual sites. This is
achieved by ramping the axial bias field of the IP trap
simultaneously with the gradient ramps.}
\bibitem[{per()}]{periods}
\bibinfo{note}{Comparable results were obtained in experiments performed with the lattice of period $d_2=5.1\,\mu$m.}
\bibitem[{2D_({\natexlab{a}})}]{2D_lowcontrast}
\bibinfo{note}{Another 15$\,\%$ of images show significantly reduced contrast
  or patterns which are hard to classify.}
\bibitem[{2D_({\natexlab{b}})}]{2D_braids}
\bibinfo{note}{For example, the braid patterns with two dislocation lines were
  observed only with more than four clouds.}
\end{thebibliography}
\end{document}